\documentclass[twocolumn,superscriptaddress,floatfix,showpacs]{revtex4-1}
\usepackage{graphicx}
\usepackage{rotating}
\usepackage{amsfonts,amsmath,color}
\usepackage{amssymb}
\usepackage{tabularx}

\newcommand{\bs}{{\bf {s}}}
\newcommand{\br}{{\bf {r}}}

\newcommand{\bv}{{\bf {v}}}

\newcommand{\etal}{{\it et al.}~}

\begin{document}

\title{Acceleration statistics in thermally driven superfluid turbulence}

\author{Andrew W. Baggaley}
\affiliation{
School of Mathematics and Statistics, University of Glasgow,
Glasgow, G12 8QW, UK
}
\author{Carlo F. Barenghi}
\affiliation{
Joint Quantum Centre Durham-Newcastle, 
School of Mathematics and Statistics, Newcastle University,
Newcastle upon Tyne, NE1 7RU, UK
}


\begin{abstract}
New methods of flow visualization near absolute zero
have opened the way to  directly compare 
quantum turbulence (in superfluid helium)
to classical turbulence (in ordinary fluids such as air or water)
and explore analogies and differences.
We present results of numerical simulations in which we examine the
statistics of the superfluid acceleration in 
thermal counterflow. 
We find that, unlike the velocity,
the acceleration obeys scaling laws similar to classical turbulence,
in agreement with a recent quantum turbulence experiment of La Mantia \etal
\end{abstract}

\maketitle

Turbulence, near omni-present in natural flows, presents an open 
and difficult problem.
It is typically studied, experimentally and theoretically, 
in a number of fluid media, all of which exhibit continuous velocity 
fields, e.g. water, air, electrically conducting plasma.
However turbulence can also be investigated in a different setting: 
low-temperature quantum fluids, which exhibit discrete vorticity fields. 
This quantum turbulence was first studied by Vinen in superfluid 
helium-4 \cite{Vinen1, Vinen2, Vinen3, Vinen4}; later studies have extended 
it to superfluid helium-3 \cite{Bradely08} 
and atomic Bose-Einstein condensates \cite{Henn09}. 
The motion of quantum fluids is strongly constrained by
quantum mechanics; notably the vorticity is concentrated in
discrete vortex filaments of fixed circulation $\kappa$ 
whose cores have atomic thickness $a_0$.
As first envisaged by Feynman, quantum turbulence consists of 
a tangle of interacting, reconnecting vortex lines.
 
In helium-4, 
quantum turbulence can readily be generated in the laboratory, 
either driving the fluid mechanically, or thermally through 
an applied heat flux;  in this article we shall focus on the latter 
method, 
which can be easily described using Landau's two--fluid
theory \cite{Donnelly}. 
A prototypical experiment consists of a channel which is
closed at one end and open to the helium bath at the other end. At the
closed end, a resistor inputs a steady flux of heat, $\dot Q$,
 into the channel. 
The heat is carried away from the resistor towards the bath 
by the normal fluid component, whereas the
superfluid component flows towards the resistor to maintain the
total mass flux equal to zero. If the relative velocity
of superfluid and normal fluid is larger than a small critical value,
the laminar counterflow of the two fluids breaks down and
a tangle of vortex lines appears, thus limiting the heat conducting
properties of helium-4.

Recent experiments have made dramatic progress in the ability to 
visualize the turbulent flow of liquid helium using tracer particles.
For example, Bewley et al. \cite{Bewley}
detected  reconnections of individual vortex lines.
Paoletti \etal \cite{paoletti2008velocity} 
discovered that in quantum turbulence the velocity statistics are
non-Gaussian, in contrast to experimental and numerical studies of 
classical turbulence which display Gaussian statistics. 
Follow--up  studies argued that this non--classical
effect  arises from the singular nature of the superfluid
vorticity \cite{White10,Baggaley-stats}.

Another important one--point observable is the distribution of 
turbulent accelerations.
In classical turbulence, Mordant et al. \cite{Mordant2004} found
that the acceleration obeys log--normal distributions; they also observed
a strong dependence of acceleration on velocity which disagrees
with the assumption of local homogeneity \cite{Mordant2005}. 
In quantum turbulence, accelerations were measured only recently by
La Mantia \etal \cite{LaMantia}. They used tracer particles
to extract Lagrangian velocity and acceleration statistics from 
thermally driven quantum turbulence at a range of temperatures 
and counterflow velocities. Their results were  striking: 
whilst observing the (now familiar) power--law nature of the 
one-point velocity statistics, their
 probability density functions (PDFs) 
of the acceleration statistics were surprisingly similar to 
classical results. 

The physics of the interactions between tracers, superfluid and normal fluid
components is complex \cite{Kivotides2008}, and
what was observed by La Mantia is only
the motion of tracers, not of the superfluid itself.
To make further progress in this problem here
we present superfluid acceleration statistics obtained by direct numerical 
simulations of thermally driven superfluid turbulence.

We model vortex lines \cite{Schwarz1985}
as oriented space curves $\bs(\xi,t)$ of infinitesimal 
thickness, where $\xi$ is arc length and $t$ is time. 
This vortex filament approach
is justified by the large separation of scales between $a_0$ and
the typical distance between vortices, $\ell$.
The governing equation of motion is
Schwarz's equation

\begin{equation}
\frac{d{\bf s}}{dt}=\bv_s+\alpha \bs' \times (\bv_n-\bv_s)
-\alpha' \bs' \times \left(\bs' \times \left(\bv_n-\bv_s\right)\right),
\label{eq:Schwarz}
\end{equation}

\noindent
where $t$ is time, $\alpha$ and $\alpha'$ are 
temperature--dependent friction coefficients 
\cite{Donnelly1998}, $\bs'=d\bs/d\xi$ is the unit
tangent vector at the point $\bs$, $\xi$ is arc length, and
$\bv_n$ is the normal fluid velocity at the point $\bf s$.
We work at temperatures comparable to La Mantia's experiment
\cite{LaMantia}; the relevant friction
coefficients are \cite{Donnelly1998}
$\alpha=0.111$, $\alpha'=0.0144$ at $T=1.65~\rm K$,
$\alpha=0.142$, $\alpha'=0.0100$ at $T=1.75~\rm K$, and
$\alpha=0.181$, $\alpha'=0.0074$ at $T=1.65~\rm K$.
The superfluid velocity $\bv_s=\bv_s^{ext}+\bv_s^{si}$ 
contains two parts: the superflow induced by the heater, $\bv_s^{ext}$,
and the self-induced velocity
of the vortex line at the point $\bs$,
given by the Biot-Savart law \cite{saffman1995vortex}
\begin{equation}
\bv_s^{si} (\bs,\,t)=
-\frac{\kappa}{4 \pi} \oint_{\cal L} \frac{(\bs-\br) }
{\vert \bs - \br \vert^3}
\times {\bf d}\br,
\label{eq:BS}
\end{equation}
where $\cal L$ is the total vortex configuration.

The techniques to discretize vortex lines
into a variable number of points 
$\bs_i$ ($i=1,\ldots, N$) held
at minimum separation $\delta/2$,
time-step Eq.~(\ref{eq:Schwarz}), de--singularize  the
Biot-Savart integrals Eq.~(\ref{eq:BS}) and evaluate them 
via a tree-method (with critical opening angle $0.3$) 
are described in a previous paper \cite{Baggaley_tree2012}.
Unlike the microscopic Gross-Pitaevskii model, 
in the vortex filament approach
vortex reconnections must be modelled algorithmically.
The reconnection algorithm used here is described in 
\cite{BaggaleyRecon} and compared to other algorithms in the
literature. All numerical simulations are performed in a periodic cube
of size ${\cal D}=0.1\,$cm. 
We take $\delta=1.6 \times 10^{-3}\,$cm and use 
time-step of $\Delta t=10^{-4}\,$s comparable to the simulations of 
Adachi \etal \cite{Adachi2010}. 
The normal fluid velocity, $\bv_n=\bv_n^{ext}$, driven by the
heater, is a prescribed 
constant flow in the positive $x$ direction. 
Our simulations are performed in the reference frame of the
superflow.  We ignore
potentially interesting physics 
arising from  boundaries, and
any influence of the quantized vortices on the normal fluid, 
but our model is sufficient for a first
study of superfluid acceleration statistics.

We present the results of five numerical simulations of counterflow
turbulence, three simulations with $v_{ns}=1\,$cm/s 
at temperatures T=1.65, 1.75, 1.85K, and two simulations for
T=1.75K at $v_{ns}=0.8\, , \,  1.2 \,$cm/s. This choice of parameters 
is motivated by the work of La Mantia \cite{LaMantia}, 
but we do not seek direct quantitative comparison with experiments,
due to the approximations inherent in our numerical approach
and in the measurements (which we discuss later),
as well as computational restrictions on the vortex line density
that can be simulated.

\begin{figure}
\begin{center}
    \includegraphics[width=0.4\textwidth]{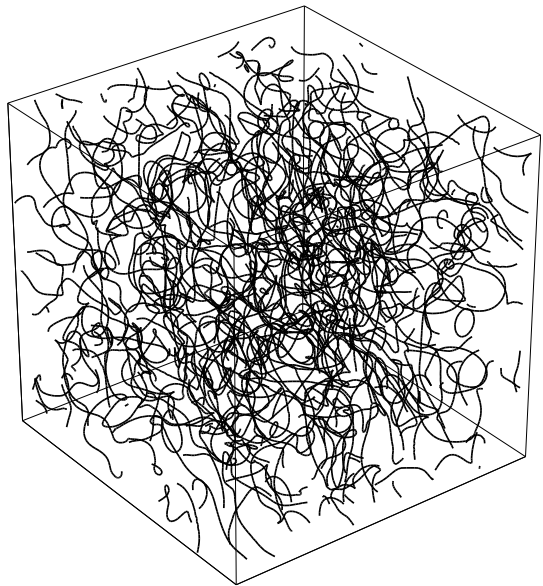} 
\caption{A snapshot of the vortex configuration 
(plotted as black space curves) at T=1.75\,K, $v_{ns}=1~\rm cm/s$,
during the quasi-steady state regime. 
Vortex line density $L=15750~\rm cm^{-2}$, 
estimated intervortex distance $\ell \approx 0.008~\rm cm$.}
\label{fig:tangle}
\end{center}
\end{figure}

All simulations are initiated with a random configuration of vortex rings 
which seed the turbulence.
As with previous studies, after and initial
transient, the vortex line density $L=\Lambda/V$ 
(defined as the superfluid vortex
length $\Lambda=\int_{\cal L} d\xi$ in the volume $V={\cal D}^3$)
saturates to a quasi-steady state (independent of the
initial seed) such that energy input from
the driving normal fluid is balanced by dissipation due to friction
and vortex reconnections. 
The intervortex distance is estimated as $\ell \approx L^{-1/2}$.
A typical vortex tangle is displayed in Fig.~\ref{fig:tangle}. 
Within the saturated regime we compute 
velocity and acceleration statistics, 
using stored velocity information at the discretization points 
$\bs_i$ via
a fourth-order upwind finite--difference scheme
\begin{equation}
\mathbf{a}_i^n=\dfrac{-\bv_i^{n-3}+6\bv_i^{n-2}-18\bv_i^{n-1}+10\bv_i^{n}+3\bv_i^{n+1}}{\Delta t},
\end{equation}
 where $\mathbf{a}_i^n$ is the acceleration of the $i^{\rm th}$ 
vortex point at the $n^{\rm th}$ time step and $\bv_i^n=d\bs_i^n/dt$ 
is the velocity of the $i^{\rm th}$ vortex point at the $n^{\rm th}$ 
time step, computed using Eq.~(\ref{eq:Schwarz}).
What we measure thus represents the Lagrangian
acceleration of ideal point tracers which are
trapped in vortex lines (hence are affected by friction), but
are unaffected by Stokes drag.

\begin{figure}
\begin{center}
    \includegraphics[width=0.4\textwidth]{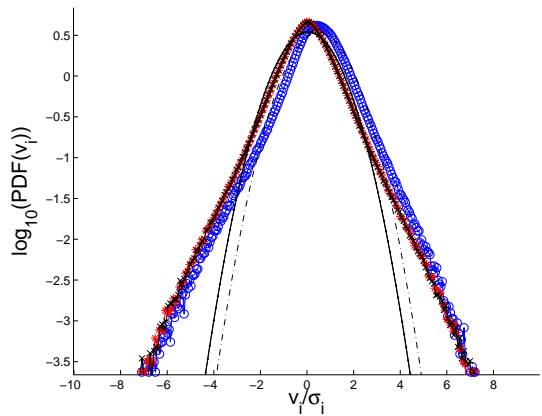} 
\caption{(Color online) Probability density functions (PDF) of turbulent
velocity components $v_i$ ($i=x,y,z)$ vs $v_i/\sigma_i$
computed from the velocity of the vortex points $d\bs_i/dt$ 
from the simulation corresponding to Fig.~\ref{fig:tangle}
($T=1.75~\rm K$, $v_{ns}=1~\rm cm/s$). 
(Blue) circles,
(red) asterisks and (black) crosses refer respectively to $i=x$, 
$i=y$ and $i=z$ components.
Gaussian fits, ${\rm gPDF}(v_i)=\frac{1}{\sqrt{2 \pi \sigma^2}}
{\rm exp}(-(v_i-\mu)^2/(2 \sigma^2)) $ for each component ($i=x$: 
dot-dashed line; $i=y$: solid line, $i=z$: solid points) 
are plotted to emphasize the deviation from Gaussianity.
Here $\sigma_x=0.1162$, $\sigma_y=0.1148$, and $\sigma_z=0.1156$.}
\label{fig:velstat}
\end{center}
\end{figure}

First we consider the velocity. PDFs of the velocity components
$v_x$, $v_y$ and $v_z$ of $\bv_i$ from
the simulation at T=1.75~K, $v_{ns}=1\,$cm/s, which are plotted 
in Fig.~\ref{fig:velstat}.
Note the power--law behavior of the tails. 
Best--fits to the data give PDF$(v) \propto v^{-3.2}$; comparable
results are obtained at different $T$ and $v_{ns}$.
The PDF's exponents, close to $-3$, are the tell--tale 
signature of quantum turbulence, and can be understood if we consider
an isolated  straight vortex line (the effect of adding the contributions
of many vortices is discussed in ref.~\cite{White10}). 
The argument is the following. 
At the distance $r$ from its axis, the vortex line
induces a velocity field $v \propto 1/r$. 
The probability $P(v)dv$
of finding the value $v$ is thus proportional to the area
$2 \pi r dr$ of the annulus between $r$ and $r+dr$; therefore
$P(v)dv \sim r dr \sim (1/v)(dv/v^2) \sim v^{-3} dv$, hence
${\rm PDF}(v) \sim v^{-3}$, in agreement
with experiments \cite{paoletti2008velocity} and numerical 
studies \cite{Baggaley-stats}.

\begin{figure}
\begin{center}
    \includegraphics[width=0.4\textwidth]{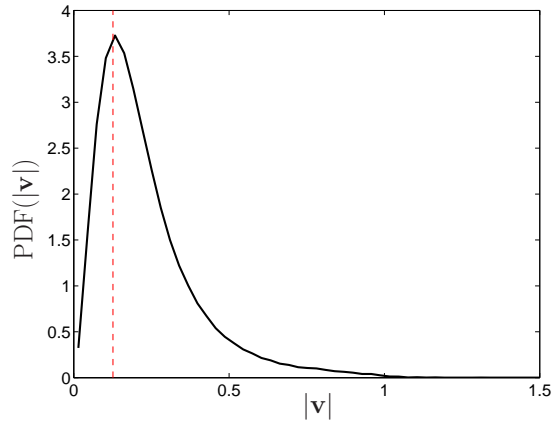} 
\caption{The probability density function (PDF) of the modulus of 
the velocity $|\mathbf{v}|$; the dashed (red) lines represents the 
`characteristic' velocity, $\kappa/\ell$. 
The PDF is computed from the data in Fig.~\ref{fig:velstat}
($T=1.75~\rm K$, $v_{ns}=1~\rm cm/s$).}
\label{fig:modv}
\end{center}
\end{figure}

It is also instructive to examine $|\bv_i|$, the modulus 
of the velocity $\bv_i$.
Numerical experiments \cite{Sherwin2012} confirm the
heuristic argument\cite{Vinen-Niemela} that 
counterflow turbulence is featureless (compared with classical turbulence), 
and the vortex tangle is characterized by the single length scale $\ell$.
The prominent peak of PDF($|\bv|$) displayed in Fig.~\ref{fig:modv} 
corresponds to the velocity scale $\kappa / \ell \approx 0.13~\rm cm/s$, 
lending further weight to the argument. The mean of the distribution,
$\langle | \mathbf{v} | \rangle =0.23~\rm cm/s$, 
is close to
the characteristic velocity of a vortex line rotating around
another line, $v_{\ell} = \kappa/(\ell/2) \approx 0.25~\rm cm/s$.

\begin{figure*}
\begin{center}
    \includegraphics[width=0.49\textwidth]{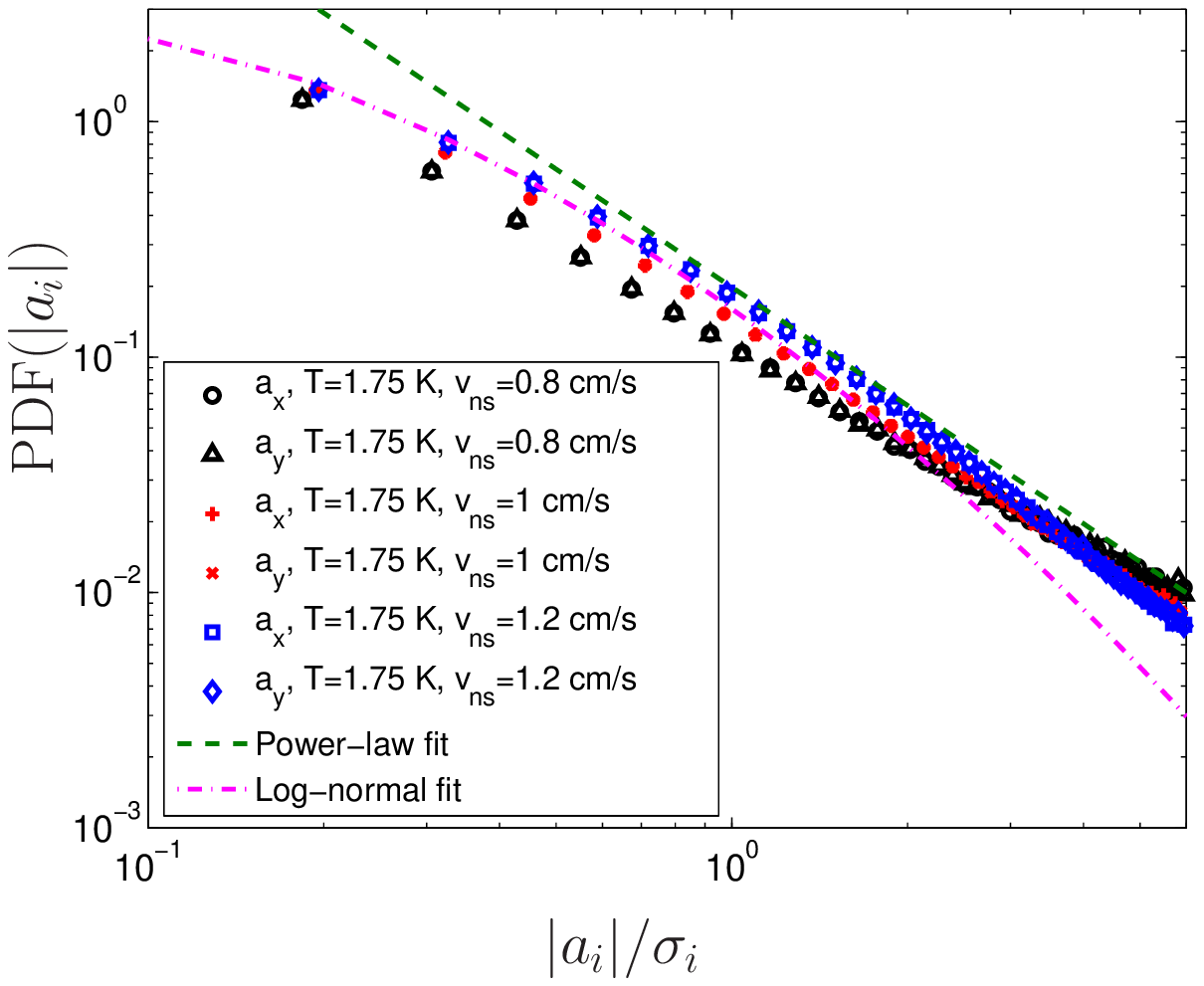} 
    \includegraphics[width=0.49\textwidth]{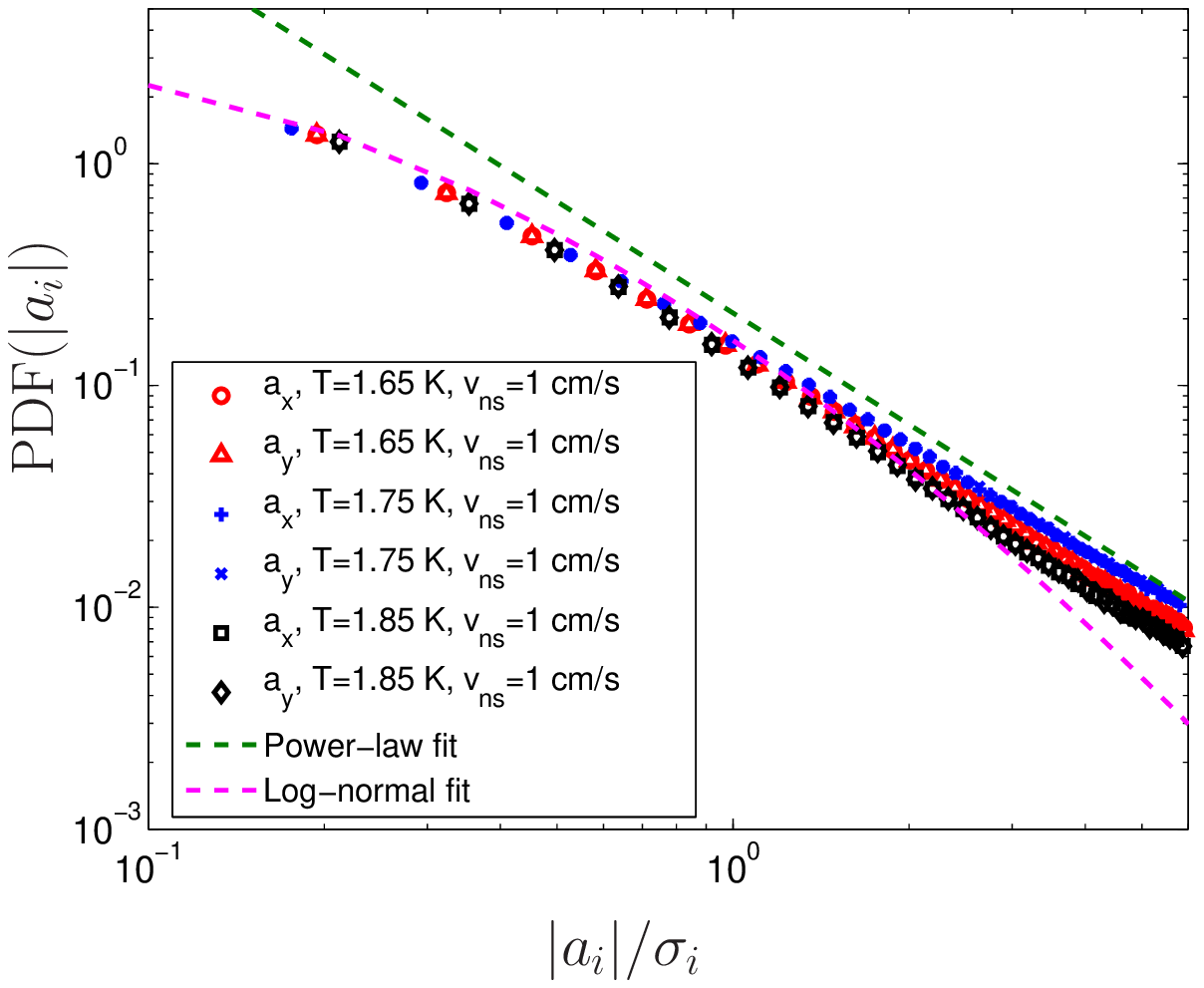} 
\caption{PDFs of the $x$ and $y$ components of the acceleration 
of the quantized vortices, scaled by the standard deviation of the 
relevant component. Left, at fixed temperature, with increasing 
counterflow velocity, right, varying the temperature with a fixed 
counterflow velocity. Log normal and power law (PDF$\sim a_i^{-5/3}$) fits to the data are plotted. 
Power law fits to the data yield 
PDF$(|a_i|/\sigma_i) \sim (a_i/\sigma_i)^\beta$ ; 
$\beta=-1.48$, T=1.75\, K, $v_{ns}=0.8 \,$ cm/s ; 
$\beta=-1.64$, T=1.75\, K, $v_{ns}=1 \,$ cm/s ; 
$\beta=-1.71$, T=1.75\, K, $v_{ns}=1.2 \,$ cm/s ; 
$\beta=-1.55$, T=1.65\, K, $v_{ns}=1 \,$ cm/s ; 
$\beta=-1.58$, T=1.75\, K, $v_{ns}=1 \,$ cm/s ;  
$\beta=-1.63$, T=1.85\, K, $v_{ns}=1 \,$ cm/s.
}
\label{fig:accstat}
\end{center}
\end{figure*}

We turn now the attention to the acceleration.
Fig.~\ref{fig:accstat} displays statistics of the 
modulus of the $x$ and $y$-components ($a_x$ and $a_y$) 
of the acceleration $\mathbf{a}_i$, normalized by 
the corresponding standard deviations ($\sigma_x$ and $\sigma_y$). 
The statistics for $a_z$ are indistinguishable from those of $a_y$.
This is not surprising, because $x$ is the longitudinal direction
of the counterflow, and the two transversal directions, $y$ and $z$,
are equivalent.
It is interesting to notice that the acceleration statistics are not
affected by the mild anisotropy of counterflow
(for example, at $T=1.75~\rm K$ and $v_{ns}=1~\rm cm/s$, 
the projected vortex lengths are such that
$L_x/L = 0.37, L_y/L = L_z/L = 0.54$.

The results displayed in Fig.~\ref{fig:accstat} are
computed at fixed temperature ($T=1.75~\rm K$) 
and varying counterflow velocities $v_{ns}$
(left), and at fixed counterflow velocity ($v_{ns}=1~\rm cm/s$) and varying
temperatures (right).
In either cases we can fit both a log-normal distribution to the data, and 
a power law to the tails of the PDF for large accelerations. 
If we apply the straight vortex line argument to the acceleration $a=v^2/r$,
we find that the probability of the value $a$ is 
$P(a) da \sim r dr \sim (1/a^{1/3}) (da/a^{4/3}) \sim a^{-5/3} da$, hence
we expect ${\rm PDF}(a) \sim a^{-5/3}$. 
The exponents shown in Fig.~\ref{fig:accstat} are in general more shallow than -5/3.
A possible explanation is that vortex 
reconnections increase the probability of large accelerations.
Lognormal distributions 
\footnote{The PDF of the log-normal distribution is given by PDF($x$)=$(1/x\sqrt{2\pi}\sigma)\exp\{ -(\ln x-\mu)^2/2\sigma^2\}$, where $\mu$ is the mean of the distribution and $\sigma^2$ is the variance.} 
are heavy-tailed (i.e. the tails of the distribution are not 
exponentially bounded), and show reasonable agreement with the data, as found by La Mantia \cite{LaMantia}. 
However it is clear that we observe a power-law scaling for the acceleration statistics in this study, with good agreement to the predicted $-5/3$ scaling.

We now consider the mean value $\langle | \mathbf{a} | \rangle$ of the
acceleration.  The previous  argument suggests that
the characteristic acceleration of a vortex line rotating 
around another line is of the order of
$a_{\ell}=v_{\ell}^2/(\ell/2)=8 \kappa^2/\ell^3$. La Mantia's
experiments support this estimate.
La Mantia reports that (insets of figure 1 of ref.~\cite{LaMantia})
that 
$\langle | \mathbf{a} |\rangle \approx 3.2$ and $1.9~\rm cm/s^2$
respectively at $T=1.64~K$, $\dot Q=586~\rm W/m^2$ and at
$T=1.86~K$, $\dot Q=595~\rm W/m^2$. If we relate the heat flux
to the counterflow velocity (via $v_{ns}={\dot Q}/(\rho_s S T)$ 
where $S$ and $\rho_s$ are the specific entropy and the superfluid
density), the counterflow velocity to the vortex line density
(via $L=\gamma^2 v_{ns}^2$, where $\gamma$ was calculated by
 Adachi  \cite{Adachi2010}), and the vortex line density
to the characteristic vortex distance
(via $\ell \approx L^{-1/2}$), we find
$a_{\ell} \approx 3.3$ and $1.1~\rm cm/s^2$ respectively, in
order of magnitude agreement with La Mantia's measurements of 
$\langle | \mathbf{a}| \rangle$. The estimate $a_{\ell}$ also
agrees with the numerical simulations. For example, at $T=1.75~\rm K$,
$v_{ns}=1~\rm cm/s$ we find $a_{\ell} \approx 16~\rm cm/s^2$ which
compares well with mean, median and mode of the computed distribution,
which are $72$, $35$ and $10~\rm cm/s^2$ respectively.

\begin{figure*}
\begin{center}
    \includegraphics[width=0.45\textwidth]{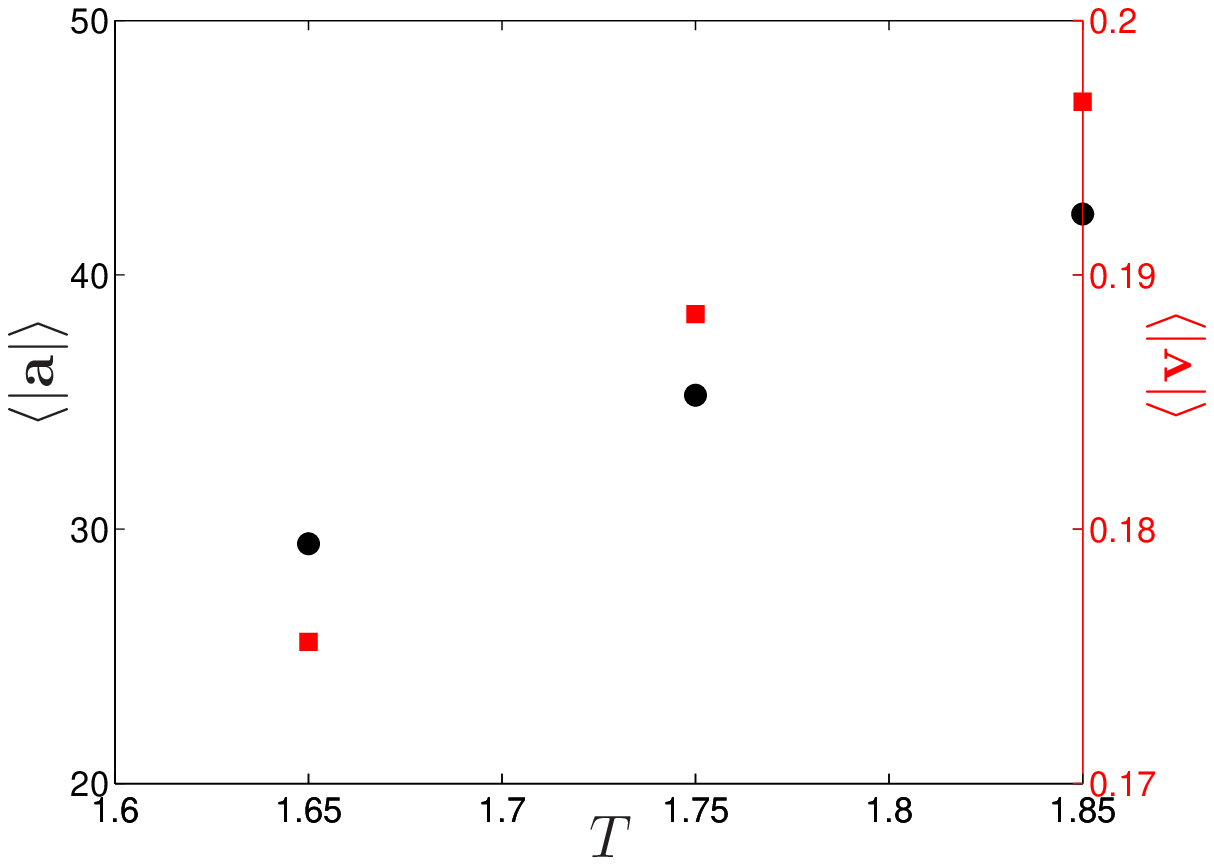} 
    \hspace{2mm}
    \includegraphics[width=0.45\textwidth]{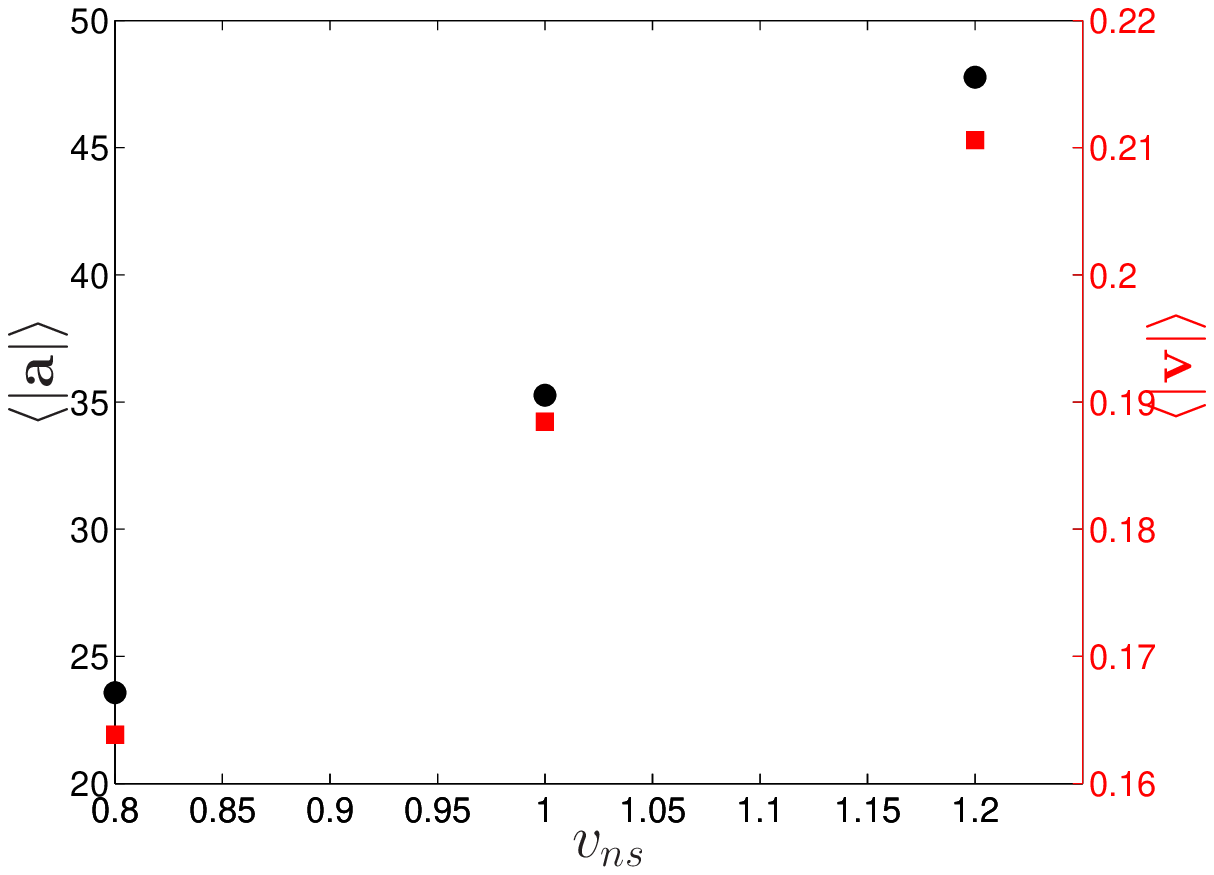} 
\caption{
Dependence of median velocity, $\langle{|\mathbf{v}|}\rangle$ ($\rm cm/s$), 
and acceleration, $\langle{|\mathbf{a}|}\rangle$ ($\rm cm/s^2$), 
on temperature $T$ (left, in $\rm K$) 
and counterflow velocity $v_{ns}$ (right, in $\rm cm/s$).
In each panel, the left axis and (black) circles correspond 
to acceleration, and the right axis and (red) squares
correspond to velocity.}
\label{fig:mean_a_v}
\end{center}
\end{figure*}

Finally, Fig.~\ref{fig:mean_a_v} shows that both
velocity and acceleration
increase with temperature $T$ (at fixed counterflow velocity $v_{ns}$)
and with $v_{ns}$ (at fixed $T$). 
La Mantia reports that $\langle | \mathbf{a} | \rangle$ increases with 
heat flux $\dot Q$ (at fixed $T$), but decreases with $T$ (at fixed $\dot Q$).  
There is no disagreement between La Mantia's results and ours.
In fact, on one hand we can write
$a_{\ell}=8 \kappa^2/\ell^2= 8 \kappa^2 \gamma^2 v_{ns}^3$: this relation
and the fact that $\gamma$ increases with increasing $T$
\cite{Adachi2010}, explains why in the
numerical simulations $\langle |\mathbf{a} | \rangle$ increases with $v_{ns}$
(at fixed $T$) and increases with $T$ (at fixed $v_{ns}$). 
On the other hand we can also write
$a_{\ell}=8 \kappa^2 (\gamma {\dot Q}/(\rho S T))^3$: this relation
accounts for La Mantia's observations that $\langle | \mathbf{a}| \rangle$ 
increases with $\dot Q$ (at fixed $T$) but decreases with $T$ 
(at fixed $\dot Q$) because the quantity $\gamma/(\rho_s S T)$ 
decreases with increasing $T$ \cite{Donnelly1998,Adachi2010}.

Our results shed light onto the complex dynamics of tracer particles.
Consider  a particle of radius $a_p$, velocity
$\bv_p$ and density $\rho_p$ which is not trapped into vortices
and moves in helium~II.
Assuming a steady uniform normal fluid, its acceleration is due to 
Stokes drag and inertial effects \cite{Poole}:

\begin{equation}
\frac{d \bv_p}{dt} = 
\frac{9 \mu_n (\bv_n-\bv_p)} {2 \rho_0 a_p^2}+
\frac{3\rho_s}{2 \rho_0} \frac{D \bv_s}{Dt},
\label{eq:Poole}
\end{equation}

\noindent
where 
$\rho$ and $\mu_n$ are helium's density and viscosity, 
and $\rho_0=\rho_p+\rho/2$.
The Stokes drag (which pulls the particle
along the normal fluid) has magnitude of the order of
$9 \beta \mu_n v_n/(2 \rho_0 a_p^2)$ where $v_n=\rho_s v_{ns}/\rho$,
$\beta v_n$ is the average slip velocity and $0<\beta<1$; 
unfortunately we do not know $\beta$ and we cannot predict the relative
importance of the two contributions to $d\bv_p/dt$.
Temporal variations of $\bv_s$ become important only after the particle
has collided with a vortex and triggered Kelvin waves \cite{Kivotides2008},
hence, for a free particle,
 the inertial term (which pulls the particle towards the nearest vortex,
effectively a radial pressure gradient) becomes
$D \bv_s /DT =\partial \bv_s \partial t + (\bv_s \cdot \nabla) \bv_s
\approx (\bv_s \cdot \nabla) \bv_s$; its magnitude is of the order of
$v_s^2/(\ell/2) =  a_{\ell}$ (which we interpreted
as the acceleration of a particle trapped into a vortex which rotates around
another vortex).  In La Mantia's experiment $\rho_0 \approx 1.9 \rho$, so the
prefactor in front of the inertial term is of order unity. The order of
magnitude agreement between the observed acceleration and our
estimate $a_{\ell}$ suggests that the Stokes term is less important 
than the inertia term. We can now interpret $a_{\ell}$
as either the acceleration of a particle trapped into a vortex which 
rotates around another vortex, of the fluctuating pressure grandient 
which attracts a free particle to a vortex line.
 
In conclusion, we have numerically
determined the one--point superfluid acceleration
statistics in counterflow turbulence, and demonstrated how mean velocity
and acceleration scale with counterflow velocity and temperature.
The importance of our results springs from the fact that
La Mantia did not measure directly the
superfluid acceleration or the vortex acceleration, but rather
the acceleration of micron--sized solid hydrogen particles, whose
dynamics is complex \cite{Poole,Kivotides2008}.
The good agreement between our findings and La Mantia's in terms of
acceleration statistics means that this difference is not crucial.
We also argue that the probability density function of one--point acceleration
statistics should follow a power law distribution, with a $-5/3$ exponent. Our
numerical results support these arguments.

The results reported by La Mantia did not distinguish
between particles which are trapped in vortices (hence move along the
imposed superflow) and particles which are free (hence move along the
normal fluid). Separate analysis of acceleration statistics of these
two groups of particles will be useful.  Theoretically,
an approach which accounts reasonably well for
velocity and acceleration statistics in classical turbulence
is the multifractal formalism \cite{Biferale2004}, which in 
principle could be adapted to model quantum turbulence.

C.F.B. is grateful to the EPSRC for grant number EP/I019413/1.

\bibliographystyle{pf}
\bibliography{biblio}

\begin{thebibliography}{10}

\bibitem{Vinen1}
W.~F. Vinen,
\newblock Proceedings of the Royal Society of London. Series A. Mathematical
  and Physical Sciences {\bf 240}, 114 (1957).

\bibitem{Vinen2}
W.~F. Vinen,
\newblock Proceedings of the Royal Society of London. Series A. Mathematical
  and Physical Sciences {\bf 240}, 128 (1957).

\bibitem{Vinen3}
W.~F. Vinen,
\newblock Proceedings of the Royal Society of London. Series A. Mathematical
  and Physical Sciences {\bf 242}, 493 (1957).

\bibitem{Vinen4}
W.~F. Vinen,
\newblock Proceedings of the Royal Society of London. Series A. Mathematical
  and Physical Sciences {\bf 243}, 400 (1958).

\bibitem{Bradely08}
D.~I. Bradley, S.~N. Fisher, A.~M. Gu\'enault, R.~P. Haley, S.~O'Sullivan,
  G.~R. Pickett, and V.~Tsepelin,
\newblock Phys. Rev. Lett. {\bf 101}, 065302 (2008).

\bibitem{Henn09}
E.~A.~L. Henn, J.~A. Seman, G.~Roati, K.~M.~F. Magalh\~aes, and V.~S. Bagnato,
\newblock Phys. Rev. Lett. {\bf 103}, 045301 (2009).

\bibitem{Donnelly}
R.~Donnelly,
\newblock {\em Quantized Vortices in Helium II},
\newblock Cambridge University Press, 1991.

\bibitem{Bewley}
G.~Bewley, M.~Paoletti, S.~Sreenivasan, and D.~Lathrop,
\newblock PNAS {\bf 105}, 13707 (2008).

\bibitem{paoletti2008velocity}
M.~S. Paoletti, M.~E. Fisher, K.~R. Sreenivasan, and D.~P. Lathrop,
\newblock Phys. Rev. Lett. {\bf 101}, 154501 (2008).

\bibitem{White10}
A.~C. White, C.~F. Barenghi, N.~P. Proukakis, A.~J. Youd, and D.~H. Wacks,
\newblock Phys. Rev. Lett. {\bf 104}, 075301 (2010).

\bibitem{Baggaley-stats}
A.~Baggaley and C.~Barenghi,
\newblock Phys. Rev. B {\bf 84}, 067301 (2011).

\bibitem{Mordant2004}
N.~Mordant, A.~M. Crawford, and E.~Bodenschatz,
\newblock Phys. Rev. Lett. {\bf 93}, 214501 (2004).

\bibitem{Mordant2005}
A.~M. Crawford, N.~Mordant, and E.~Bodenschatz,
\newblock Phys. Rev. Lett. {\bf 94}, 024501 (2005).

\bibitem{LaMantia}
M.~La~Mantia, D.~Duda, M.~Rotter, and L.~Skrbek,
\newblock Journal of Fluid Mechanics {\bf 717} (2013).

\bibitem{Kivotides2008}
D.~Kivotides, C.~Barenghi, and Y.~Sergeev,
\newblock Physical Review B {\bf 77}, 014527 (2008).

\bibitem{Schwarz1985}
K.~W. Schwarz,
\newblock Phys. Rev. B {\bf 31}, 5782 (1985).

\bibitem{Donnelly1998}
R.~J. Donnelly and C.~F. Barenghi,
\newblock J. Phys. Chem. Ref. Data {\bf 27}, 1217 (1998).

\bibitem{saffman1995vortex}
P.~Saffman,
\newblock {\em Vortex Dynamics},
\newblock Cambridge Monographs on Mechanics, Cambridge University Press, 1995.

\bibitem{Baggaley_tree2012}
A.~Baggaley and C.~Barenghi,
\newblock Journal of Low Temperature Physics {\bf 166}, 3 (2012).

\bibitem{BaggaleyRecon}
A.~Baggaley,
\newblock J. Low Temp. Phys. {\bf 168}, 18 (2012).

\bibitem{Adachi2010}
H.~Adachi, S.~Fujiyama, and M.~Tsubota,
\newblock Phys. Rev. B {\bf 81}, 104511 (2010).

\bibitem{Sherwin2012}
A.~W. Baggaley, L.~K. Sherwin, C.~F. Barenghi, and Y.~A. Sergeev,
\newblock Phys. Rev. B {\bf 86}, 104501 (2012).

\bibitem{Vinen-Niemela}
W.~Vinen and J.~Niemela,
\newblock J. Low Temp. Physics {\bf 128}, 167 (2000).

\bibitem{Note1}
The PDF of the log-normal distribution is given by PDF($x$)=$(1/x\protect \sqrt
  {2\pi }\sigma )\protect \qopname \relax o{exp}\protect \{ -(\protect \qopname
  \relax o{ln}x-\mu )^2/2\sigma ^2\protect \}$, where $\mu $ is the mean of the
  distribution and $\sigma ^2$ is the variance.

\bibitem{Poole}
D.~R. Poole, C.~F. Barenghi, Y.~A. Sergeev, and W.~F. Vinen,
\newblock Phys. Rev. B {\bf 71}, 064514 (2005).

\bibitem{Biferale2004}
L.~Biferale, G.~Boffetta, A.~Celani, B.~J. Devenish, A.~Lanotte, and F.~Toschi,
\newblock Phys. Rev. Lett. {\bf 93}, 064502 (2004).

\end{thebibliography}
\end{document}